\pdfoutput=1
\documentclass[superscriptaddress, 12pt, aps, prl,  preprint, longbibliography]{revtex4-1}
\usepackage{graphicx}
\usepackage{epstopdf}
\usepackage{color}
\usepackage[utf8]{inputenc}
\usepackage{amsmath, amssymb, physics}

\begin{document}

\title{Raman and IR spectra of water under graphene nanoconfinement at ambient and extreme pressure-temperature conditions: a first-principles study}
\author{Rui Hou}
\altaffiliation{These authors contributed equally to this work.}
\affiliation{Department of Physics, Hong Kong University of Science and Technology, Hong Kong, China}
\affiliation{HKUST Shenzhen-Hong Kong Collaborative Innovation Research Institute, Shenzhen, China}
\author{Chu Li}
\altaffiliation{These authors contributed equally to this work.}
\affiliation{Department of Physics, Hong Kong University of Science and Technology, Hong Kong, China}
\affiliation{HKUST Shenzhen-Hong Kong Collaborative Innovation Research Institute, Shenzhen, China}
\author{Ding Pan}
\email{dingpan@ust.hk}
\affiliation{Department of Physics, Hong Kong University of Science and Technology, Hong Kong, China}
\affiliation{Department of Chemistry, Hong Kong University of Science and Technology, Hong Kong, China}
\affiliation{HKUST Shenzhen-Hong Kong Collaborative Innovation Research Institute, Shenzhen, China}

\date{\today}

\begin{abstract}
The nanoconfinement of water can result in dramatic differences in its physical and chemical properties compared to bulk water. However, a detailed
molecular-level understanding of these properties is still lacking.
Vibrational spectroscopy, such as Raman and infrared, is a popular experimental tool for studying the structure and dynamics of water, and is often complemented by atomistic simulations to interpret experimental spectra, but there have been few theoretical spectroscopy studies of nanoconfined water using first-principles methods at ambient conditions, let alone under extreme pressure-temperature conditions.
Here, we computed the Raman and IR spectra of water nanoconfined by graphene at ambient and extreme pressure-temperature conditions using ab intio simulations. 
Our results revealed alterations in the Raman stretching and low-frequency bands due to the graphene confinement. 
We also found spectroscopic evidence indicating that nanoconfinement considerably changes the tetrahedral hydrogen bond network, which is typically found in bulk water. 
Furthermore, we observed an unusual bending
band in the Raman spectrum at $\sim$10 GPa and 1000 K, which is attributed to the unique molecular structure of confined ionic water. 
Additionally, we found that at $\sim$20 GPa and 1000 K, confined water transformed into a superionic fluid, making it challenging to identify the IR stretching band.
Finally, we computed the ionic conductivity of confined water in the ionic and superionic phases.
Our results highlight the efficacy of Raman and IR spectroscopy in studying the structure and
dynamics of nanoconfined water in a large pressure-temperature range.
Our predicted Raman and IR spectra can serve as a valuable guide for future experiments.
\end{abstract}

\maketitle

\section{Introduction}
Water, arguably the most ubiquitous solvent in our daily lives, exists not only in the bulk phase, but also under confinement in nature.
The confined water or aqueous solutions can be even found as deep as in Earth's mantle, where pressure (P) may reach tens of GPa and temperature (T) as high as more than 1000 K \cite{gautam2017structure, Marquardt2018Structure, stolte2022nanoconfinement}.
Recent advances in the synthesis and manipulation of low dimensional nanostructures, e.g. graphene, have enabled precise control and investigation of the behaviors of nanoconfined aqueous solutions \cite{Munoz-Santiburcio2021Confinement}. The use of slit pores with confinement widths smaller than 1 nm, for example, has become feasible \cite{lowdielectric2018, nanoconfinement2022Ruiz, Munoz-Santiburcio2021Confinement}. The strong effect of nanoconfinement can result in dramatic differences in the physical and chemical properties of confined aqueous solutions as compared to the bulk phase \cite{lowdielectric2018, lowdielectric2022, 2Dice2016,2Dice2022}. However, a detailed molecular-level understanding of these properties is still lacking.

Vibrational spectroscopy, such as Raman and infrared (IR),  is a widely used tool for investigating the structure and dynamics of water and aqueous solutions in various environments at the molecular scale \cite{auer2008ir, pan2020first}. This technique is highly sensitive to inter- and intramolecular chemical environment of water molecules, which determine the strength of the hydrogen bond  network of water. The sensitivity of vibrational spectroscopy provides a unique approach to investigating the hydrogen bond network and its dynamics, allowing for a deeper understanding of water's behavior at the molecular level \cite{spectro-review2010, spectro-review2016}. Over the past few decades, spectroscopic studies on bulk water and water at interfaces have significantly improved our understanding of water's vibrational modes and their relationship to the hydrogen bond network \cite{auer2008ir,spectro-review2010, spectro-review2016, spectro-review2020Seki}.

Vibrational spectroscopy provides an indirect means of probing molecular structures, and is often complemented by atomistic simulations to aid in the interpretation of experimental spectra \cite{auer2008ir}.
Experimental spectroscopy studies of nanoconfined water are technically more challenging than those of bulk water, and the results are also more difficult to interpret \cite{Munoz-Santiburcio2021Confinement}.
The elevated pressure and temperature conditions make experimental measurements even more complicated \cite{gygi2005ab}.
Ab initio molecular dynamics (AIMD) simulations do not rely on experimental input or empirical parameters, and can treat electronic polarizability and charge polarization, which are crucial for Raman and IR spectroscopy, at the quantum mechanical level \cite{car1985unified, galli1991ab, marx2009ab}.
Therefore, the AIMD method is widely regarded as one of the most reliable methods for interpreting experimental data and making predictions, particularly at extreme P-T conditions \cite{gygi2005ab}.
However, there have been few theoretical spectroscopy studies of nanoconfined water using the AIMD method at ambient conditions, let alone under extreme P-T conditions.

In this paper, we computed the Raman and IR spectra of water nanoconfined by graphene at ambient and extreme P-T conditions using AIMD simulations. 
We compared the spectra of confined water with those of bulk water and observed alterations in the Raman stretching band and low-frequency band due to the extreme spatial confinement. 
By analyzing the inter- and intramolecular contributions, we discussed the effect of nanoconfinement on the hydrogen bond network.
Additionally, we investigated the impact of extreme P-T conditions on the Raman and IR spectra of confined water and identified an unusual bending band in the Raman spectrum at around 10 GPa and 1000 K, resulting from the unique molecular structure of confined water.
Due to the promotion of water dissociation under nanoconfinement, the stretching band of the IR spectrum became challenging to identify at approximately 20 GPa and 1000 K, when water transformed into a superionic liquid. Finally, we computed the ionic conductivity of confined water at extreme P-T conditions.
Our results highlight the efficacy of Raman and IR spectroscopy in exploring the structural and dynamic properties of nanoconfined water at both ambient and extreme P-T conditions, and our predicted Raman and IR spectra can serve as a valuable guide for future experimental investigations.

\section{Methods}
\begin{figure}
\centering
\vspace{5mm}
\includegraphics[width=1.0 \textwidth]{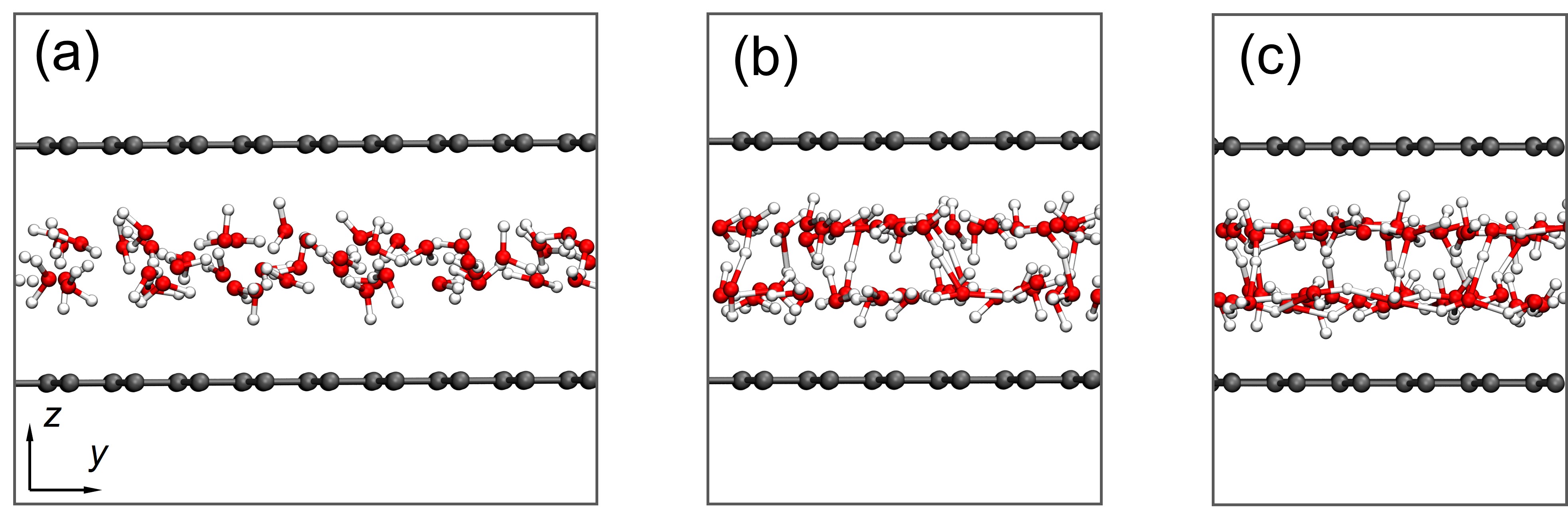}
\caption{ Snapshots of water nanoconfined by two graphene sheets. The configurations were randomly selected from ab initio molecular dynamics simulations trajectories:
(a) ambient conditions; (b)$\sim$10 GPa, 1000 K; and (c) $\sim$20 GPa, 1000 K. Water shows two bilayers in (b) and (c). The z direction is perpendicular to the graphene sheets.}
\label{snapshot}
\end{figure}

We performed AIMD simulations with the Born-Oppenheimer approximation using the Qbox code (http://qboxcode.org) \cite{Gygi2008Architecture}. The exchange-correlation (xc) functional is Perdew-Burke-Ernzerhof (PBE) \cite{Perdew1996Generalized}. 
While it is widely recognized that PBE is inadequate for describing aqueous systems at ambient conditions \cite{gillan2016perspective}, our previous studies have demonstrated that it performs better for the equation of state and dielectric properties of water \cite{Pan2013Dielectric, Pan2014Refractive}, as well as the carbon speciation in water at extreme P-T conditions 
than at ambient conditions \cite{pan2020first, stolte2022nanoconfinement}.
Furthermore, several previous spectroscopy studies of bulk water at ambient and extreme P-T conditions have utilized PBE as the xc functional \cite{zhang2010first, quantumcal2013wan, abinitio2018}. 
Although PBE underestimates the O-H stretching frequencies and overestimates the width of the O-H stretching band, it can correctly produce the main features of experimental vibrational spectra \cite{quantumcal2013wan}.
For consistency and ease of comparison, we chose not to employ more advanced functionals in this study. We mainly focus on the change of vibrational spectra due to the extreme spatial confinement, which is less affected by the choice of xc functionals, rather than accurately reproducing experimental spectra.
Because the semilocal PBE xc functional lacks the dispersion interaction, we modeled the graphene sheets using a distance-dependent potential acting on the oxygen atoms, which was calibrated to the interaction energies obtained from diffusion quantum Monte Carlo calculations \cite{Brandenburg2019Physisorption, stolte2022nanoconfinement}. The distance between two graphene sheets is 0.77 nm (see Fig. \ref{snapshot}). We used deuterium instead of hydrogen to enable a larger AIMD time step, but still refer to these atoms as hydrogen atoms. More simulation details are given in the supporting information. 

The Raman spectra were computed on the fly using the auto-correlation functions of electronic polarizabilities, which were calculated every 20 MD steps along trajectories with a length between 50 and 100 picoseconds. The polarizability tensors were calculated using the finite field method \cite{PhysRevLett.89.117602}, and the magnitude of the macroscopic electric field is 0.001 Hartree/Bohr. The electronic dipole was defined using the centers of charge of maximally localized Wannier functions (MLWFs) \cite{marzari2012maximally} with the refinement correction proposed by Stengel and Spaldin \cite{stengel2006accurate}. For one H$_2$O, OH$^-$, H$_3$O$^+$, or O$^{2-}$  molecule/ion, its electronic dipole moment is calculated as  $\pmb{\mu}^{elec}_i = -2e \sum_{j=1}^{4} \vb{R}_j^{MLWF}$, where $e$ is the elementary charge, and $\vb{R}_j^{MLWF}$ is the position vector pointing from the oxygen atom to the center of charge of the $j$th MLWF with the minimum-image convention. There are four MLWFs closely around each oxygen atom in our simulations. The effective electronic polarizability tensor of the $i$th molecule or ion is $\alpha_i^{eff} = \Delta \pmb{\mu}_i^{elec}/\vb{E}$, where $\Delta \pmb{\mu}_i^{elec}$ denotes the change of the electronic dipole moment of the $i$th molecule or ion induced by the macroscopic electric field $\vb{E}$. The electonic polarizability tensor of the whole simulation box is calculated as the sum of the effective molecular or ionic polarizabilities: $\alpha = \sum_{i=1}^N \alpha_i^{eff}$, where $N$ is the number of oxygen atoms. Note that the effective polarizability of a molecule or ion is not the same as the molecular or ionic polarizability, because the macroscopic field $\vb{E}$ is not the local electric field \cite{pan2018communication}.

The isotropic and anisotropic Raman spectra were calculated separately using Eqs. (\ref{eq:6}) and (\ref{eq:7}), respectively \cite{quantumcal2013wan}:

\begin{equation}
\label{eq:6}
R_{\text{iso}}(\omega) \propto \dfrac{\hbar \omega}{k_BT} \int{}dt\text{e}^{-\text{i}\omega t }\langle \bar{\alpha}(0)\bar{\alpha}(t)\rangle
\end{equation}

\begin{equation}
\label{eq:7}
R_{\text{aniso}}(\omega) \propto \dfrac{\hbar \omega}{k_BT} \int{}dt\text{e}^{-\text{i}\omega t }\langle \dfrac{2}{15} \text{Tr}\beta (0) \beta (t)\rangle.
\end{equation}
In these equations, $\omega$ represents the frequency, $k_B$ is the Boltzmann constant, $t$ represents the correlation time, and Tr is the trace matrix operator. $\bar{\alpha}$ and $\beta$ are the isotropic and anisotropic components of the polarizability tensor $\alpha$. Specifically, $\bar{\alpha}$ is calculated as $\dfrac{1}{3}\text{Tr}\alpha$ and $\beta$ is calculated as $\alpha - \bar{\alpha}\textbf{I}$, where $\textbf{I}$ is the identity tensor. The unpolarized Raman spectra were obtained by taking a linear combination of the isotropic and anisotropic spectra using the formula: $R_{\text{unpol}} = R_{\text{iso}} + \dfrac{7}{4}R_{\text{aniso}}$. Finally, a Gaussian broadening with a full width at half maximum (FWHM) of 80 cm$^{-1}$ was used to smooth the spectra. All the calculated spectra are normalized between 0 and 3200 cm$^{-1}$.

Using the effective molecular or ionic polarizability $\alpha_i^{eff}$, we can decompose isotropic Raman spectra obtained by Eq. (\ref{eq:6}) as,
\begin{equation}
\label{Raman-decomp}
R_{\text{iso}}(\omega) \propto 
\frac{\hbar \omega}{k_BT} \int{}dt\text{e}^{-\text{i}\omega t }\langle \sum_{\substack{i=1 \\ i\neq j}}^N \sum_{j=1}^N \bar{\alpha}_{i}^{eff}(0)\bar{\alpha}_{j}^{eff}(t) + \sum_{i=1}^{N}\bar{\alpha}_{i}^{eff}(0)\bar{\alpha}_{i}^{eff}(t)\rangle
\end{equation}
where the first and second terms give the inter- and intramolecular Raman spectra, respectively. We can also decompose anisotropic Raman spectra in a similar way. 

The IR spectra were computed on the fly using the auto-correlation functions of total dipole moments \cite{zhang2010first, abinitio2018}:
\begin{equation}\label{IR-cal}
    I(\omega) \propto \frac{2\pi}{3cVk_BT}
    \int{}dt\text{e}^{-\text{i}\omega t }\langle \dot{\vb{M}}(0)\dot{\vb{M}}(t)\rangle,
\end{equation}
where $c$ is the speed of light, $V$ is the cell volume.
The total dipole moment of a simulation cell is calculated as 
\begin{equation} \label{M_pbc}
\vb{M} = \sum_{i=1}^N \pmb{\mu}^{elec}_i - 2e\sum_{i=1}^{N}\vb{R}_i^O + e\sum_{i=1}^{2N}\vb{R}_i^H
\end{equation}
where $\vb{R}_i^O$ and $\vb{R}_i^H$ are the unfolded positions of oxygen and hydrogen atoms.
Based on the modern theory of polarization, 
we cannot uniquely define the total dipole moment of ionic fluids other than a small change of $\vb{M}$ due to periodic boundary conditions \cite{resta2007theory}, so we used the time derivative $\dot{\vb{M}}$ in the IR spectra calculations.
At ambient conditions, water is a molecular liquid, 
so the total dipole moment $\vb{M}$ can be written as the sum of molecular dipole moments, which enables us to apply the method similar to Eq. (\ref{Raman-decomp}) to divide Eq. (\ref{IR-cal}) into the inter- and intamolecular contributions.

Using the definition of the total dipole in Eq. (\ref{M_pbc}), we can further calculate the ionic conductivity of water based on the Einstein relation \cite{cavazzoni1999superionic, abinitio2018}:
\begin{equation}
    \sigma = \lim_{t\rightarrow \infty} \frac{1}{6tk_BTV}\langle [\vb{M}(t)-\vb{M}(0)]^2 \rangle,
\end{equation}
which is mathematically equivalent to the Green-Kubo relation.

\section{Results and discussion}

\subsection {Raman spectra at ambient conditions}

\begin{figure}
\centering
%\vspace{5mm}
\includegraphics[width=1.0 \textwidth]{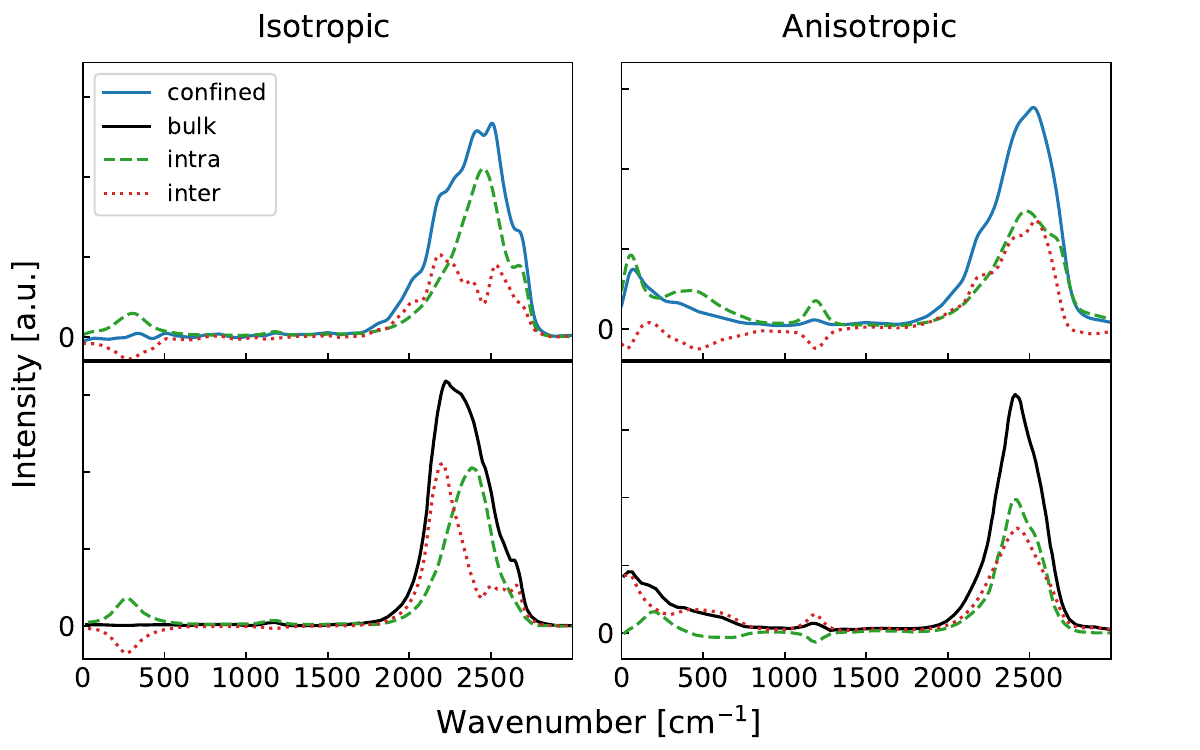}
\caption{Raman spectra of graphene-confined and bulk water at ambient conditions.
The left and right panels show the isotropic and anisotopic Raman spectra, respectively. The upper and lower panels show the Raman spectra of confined (blue lines) and bulk water (black lines), respectively. 
The intra- and intermolecular contributions are indicated by green dashed and red dotted lines, respectively. The spectra of bulk water are from Ref. \cite{quantumcal2013wan}.}
\label{ambient}
\end{figure}

Fig. \ref{ambient} shows the isotropic and anisotropic Raman spectra of graphene-confined water and bulk water at ambient conditions. 
We set the temperature at 400 K and the same pressure as that of the bulk PBE water at 400 K and 1 g/cm$^3$, consistent with the previous studies \cite{quantumcal2013wan}. 
The overall Raman spectra of graphene-confined water are similar to those of bulk water, but have some important differences. 
The main Raman band appears between 1800 and 2800 cm$^{-1}$, which is attributed to the OH stretching. The stretching band of graphene-confined water shifts towards high frequency in both isotropic and anisotropic spectra, as compared to that of bulk water, which can be explained by more dangling bonds at the graphene-water interface \cite{Munoz-Santiburcio2021Confinement}. 
We further decomposed the Raman spectra into inter- and intramolecular contributions.
In the isotropic spectrum of bulk water, 
the intramolecular contribution to the stretching band has a higher peak frequency than the intermolecular contribution, whereas for the graphene-confined water,
the intermolecular contribution splits into two main peaks, and the main peak of the intramolecular contribution is located between these two peaks.
The bending mode of water molecules is weakly observed at 1182 cm$^{-1}$, which is more apparent 
in the anisotropic Raman spectra for both graphene-confined and bulk water.

In the region below 500 cm$^{-1}$,
the isotropic Raman spectrum of confined water shows a few tiny peaks, while bulk water has no apparent features. Wan et al. found that the charge fluctuations between hydrogen-bonded water molecules in the bulk water may cause intermolecular anticorrelation \cite{quantumcal2013wan}, so the inter- and intramolecular contributions cancel each other and there are little signals in the total isotropic spectrum below 500 cm$^{-1}$. 
Confinement can significantly alter the hydrogen bond network and consequently, the charge fluctuations. As a result, small signals can be observed in the confined system, which are distinct from those reported in the bulk phase.

In Fig. \ref{ambient}, the anisotropic spectrum of confined water exhibits a pronounced peak at 71 cm$^{-1}$ in the frequency region below 300 cm$^{-1}$, whereas both first-principles calculations and experiments show two peaks at 60 and 200 cm$^{-1}$ for bulk water \cite{quantumcal2013wan}. The origin of these two peaks was controversial. 
By combining with the IR spectrum, Wan et al. figured out that they are attributed to the intra- and intermolecular contributions, respectively \cite{quantumcal2013wan, PhysRevLett.95.187401}. The 200 cm$^{-1}$ peak observed in the anisotropic spectrum of bulk water is related to intermolecular charge fluctuations that occur within the tetrahedral hydrogen bond network \cite{PhysRevLett.95.187401}. However, nanoconfinement significantly distorts the bulk hydrogen bond network,  so the 200 cm$^{-1}$ peak is absent in the anisotropic spectrum, as shown in Fig. \ref{ambient}.

\subsection{Raman spectra under extreme P-T conditions}
\begin{figure}
\centering
%\vspace{5mm}
\includegraphics[width=0.6 \textwidth]{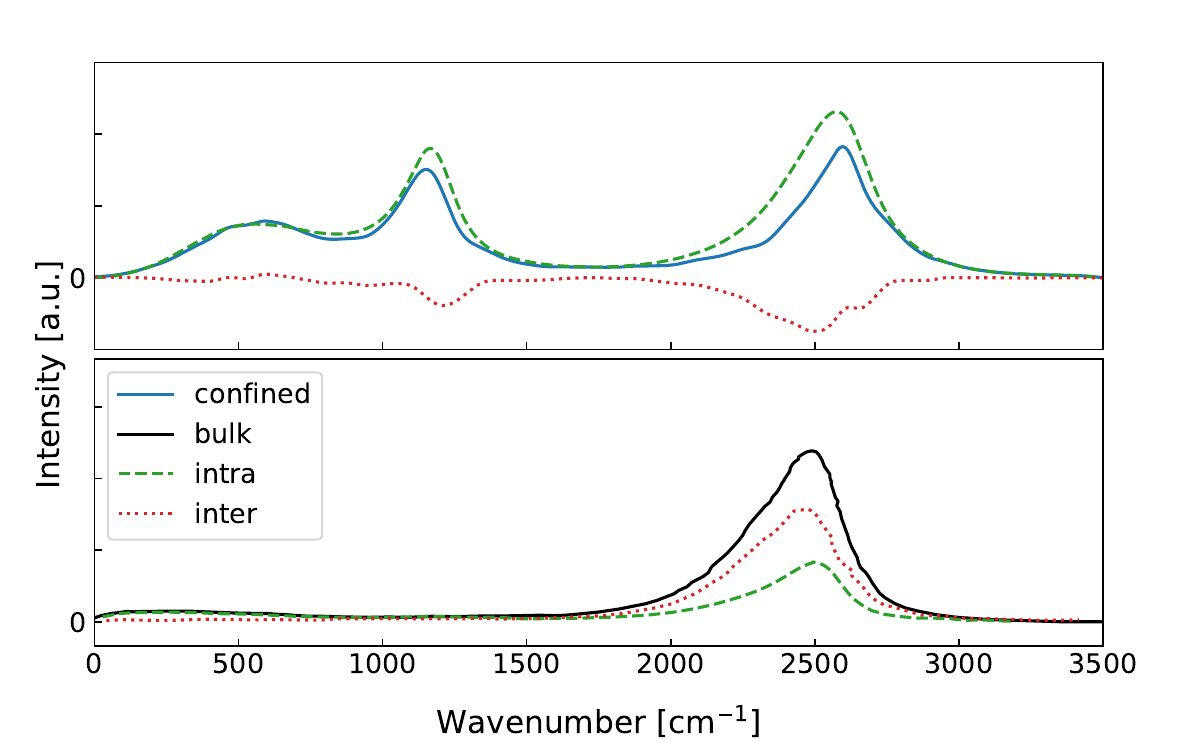}
\caption{
Raman spectra of graphene-confined and bulk water at $\sim$10 GPa and 1000 K. The upper and lower panels show the Raman spectra of confined (blue lines) and bulk water (black lines), respectively.
The intra- and intermolecular contributions are indicated by green dashed and red dotted lines, respectively. The spectra of bulk water are from Ref. \cite{abinitio2018}, and
the frequencies were scaled down by 0.735 to convert the stretching band of H$_2$O to that of D$_2$O \cite{ohto2019accessing}. 
}
\label{Raman-10GPa}
\end{figure}

\begin{figure}
\centering
\vspace{5mm}
\includegraphics[width=0.6 \textwidth]{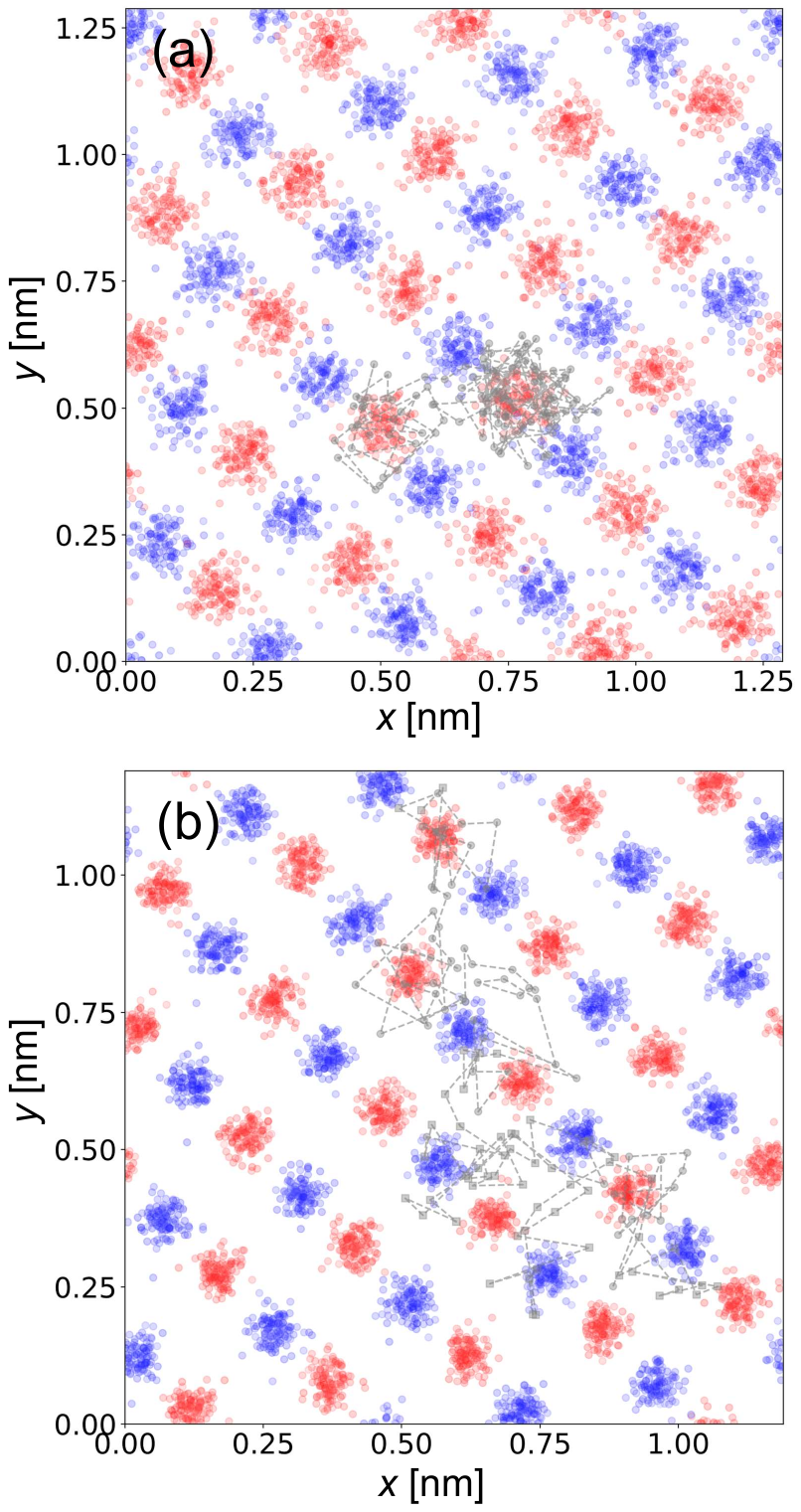}
\caption{
Positions of atoms projected to the xy plane under graphene nanoconfinement at extreme P-T conditions.
The blue and red dots represent the positions of oxygen atoms in the bottom and upper bilayers, respectively (see Fig. \ref{snapshot}).
Gray dots show the transfer trajectory of one hydrogen atom
in the bottom (circles) or upper (squres) bilayers.
Dashed lines indicate the time evolution of trajectories.
The projected atom positions were recorded every 0.242 ps along the 29 ps trajectories at
(a) $\sim$10 GPa and 1000 K, and (b)  $\sim$20 GPa and 1000 K.}
\label{projected-O-H}
\end{figure}

Fig. \ref{Raman-10GPa} shows that unpolarized Raman spectra of graphene-confined water and bulk water at $\sim$10 GPa and 1000 K. 
Experimental setups become more complex under high-P and high-T conditions, making it challenging to align incident and scattered light for polarization measurements \cite{goncharov2012raman}. Therefore, we focus on unpolarized Raman spectra obtained from our calculations at extreme conditions.
Unlike the spectra under ambient conditions, the spectrum of confined water exhibits a much more prominent bending band with a peak at 1151 cm$^{-1}$ than that of bulk water.  
In bulk water or a water molecule in the gas phase, the bending mode has a small Raman cross section due to the relatively small change in molecular polarizability.
At $\sim$10 GPa and 1000 K under nanoconfinement, the molecular polarizability of water appears to change more significantly than in the bulk phase. To better understand this phenomenon, we examined the molecular structure and dynamics of confined water.
We found that under nanoconfinement the oxygen and hydrogen atoms have different diffusion constants, 0.0542 and 0.493 nm$^2$/ns, respectively, whereas in the bulk phase, the mean squared displacements of these two atoms have similar slopes \cite{abinitio2018}, indicating that water molecules dissociate more readily under nanoconfinement than in the bulk phase, which is consistent with the previous studies at close to ambient conditions \cite{Munoz-Santiburcio2017Nanoconfinement}. 
Nanoconfined water at $\sim$10 GPa and 1000 K can be considered as an ionic liquid, where the OH bond breaks and forms frequently, while the bulk water at the same P-T condition is still a molecular liquid, despite some proton transfers. 
Fig. \ref{projected-O-H}(a) shows that the trajectories of oxygen atoms and one hydrogen atoms projected to the xy plane parallel to the graphene sheets within 29 ps. The average positions of the oxygen atoms in the two bilayers under confinement have a shape that is similar to a rhombohedral lattice (see Fig. S2 (a) and (b) in the supporting information). Note that the oxgyen atoms still diffuse slowly as shown by the mean squred displacement in Fig. S1 in the supporting information, so they are not in a solid phase. 
Fig. \ref{projected-O-H}(a)  also shows the trajectory of one randomly picked hydrogen atom. It mostly moved around two O atoms in the bottom water bilayer within 29 ps.  
Because the OH bond length varies greatly for the ionic water in the bending mode, the molecular polarizability of H$_2$O and OH$^-$
changes significantly, and consequently this mode has an unusually large Raman cross section. 

\begin{figure}
\centering
\vspace{5mm}
\includegraphics[width=0.6 \textwidth]{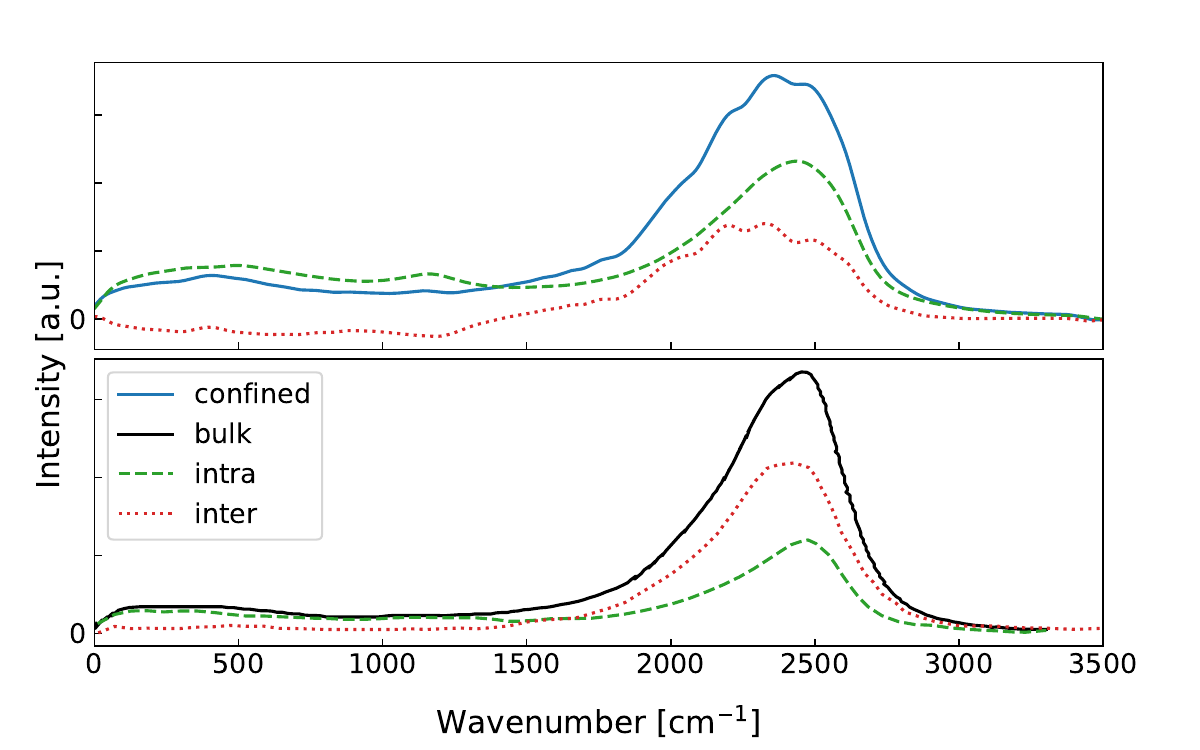}
\caption{Raman spectra of graphene-confined and bulk water at $\sim$20 GPa and 1000 K. The upper and lower panels show the Raman spectra of confined (blue lines) and bulk water (black lines), respectively.
The intra- and intermolecular contributions are indicated by green dashed and red dotted lines, respectively. The spectra of bulk water are from Ref. \cite{abinitio2018}, and
the frequencies were scaled down by 0.735 to convert the stretching band of H$_2$O to that of D$_2$O \cite{ohto2019accessing}. 
}
\label{Raman-20GPa}
\end{figure}

We further increased pressure to $\sim$20 GPa at 1000 K.
Fig. \ref{Raman-20GPa} shows the Raman spectra of graphene-confined water and bulk water. It is interesting to see that the bending band becomes very small again. 
For the graphene confined water, the diffusion constants of the hydrogen atoms, 4.41 nm$^2$/ns, is two orders of magnitude larger than that of the oxygen atoms, 0.0129 nm$^2$/ns.
Fig. S2 (c) and (d) show that the O atoms have a distorted rhombohedral lattice with a slightly enlarged lattice constant $c$, while the hydrogen atoms can diffuse in a long distance between the upper and bottom bilayers. 
We considered the confined water becomes a superionic liquid at  $\sim$20 GPa at 1000 K.
Because the hydrogen atoms are not closely bonded to the oxygen atoms, the molecular polarizability is essentially that of the O$^{2-}$ ions. The O$^{2-}$ ions strongly attract electrons, which limits the variation of polarizability. As a result, the Raman signal of the bending mode is weak.
While previous spectroscopy studies of water have predominantly focused on the OH stretching band, our understanding of the bending band remains limited \cite{spectro-review2020Seki}. However, our current work on nanoconfined ionic water shows that the intensity of the bending band can be comparable to that of the stretching band. This suggests that the bending mode could serve as a valuable tool for investigating the molecular structure of confined water.

Another interesting finding is that the intermolecular contributions of confined water to the stretching and bending Raman bands are negative at $\sim$10 GPa and 1000 K, indicating that the vibrations are anticorrelated along the hydrogen bond network under confinement.
At $\sim$20 GPa and 1000 K, the intramolecular contributions are larger than the intermolecular ones, which is the opposite of the situation at ambient conditions. The possible reason is that the hydrogen bond network under confinement is smaller than that in the bulk phase,  resulting in Raman signals that are primarily from intramolecular vibrations.

\subsection{IR spectra at ambient and extreme P-T conditions}

We also computed the IR spectra of water under confinement.
At ambient conditions, the stretching band of confined water is broader than that of bulk water.
For both confined and bulk water, 
the high frequency stretching comes more from the intramolecular contribution. 
In the low frequency region below 700 cm$^{-1}$, our spectrum is similar to the previous study by THz spectroscopy with the graphene interlayer distance of 0.668 nm \cite{nanoconfinement2022Ruiz}. This frequency region consists of two main bands: the intermolecular stretching band with a peak at $\sim$200 cm$^{-1}$ and the librational band above 500 cm$^{-1}$.
At $\sim$10 GPa and 1000 K, the stretching band of the confined water becomes broader than that in the bulk phase, and with increasing pressure to $\sim$20 GPa, the stretching band becomes even more difficult to identify.
As we observed in the Raman spectra, water transforms into an ionic liquid at $\sim$10 GPa and 1000 K, and a superionic liquid at $\sim$20 GPa and 1000 K. Therefore, under nanoconfinement, the lifetime of the OH bonds becomes shorter, resulting in weaker O-H stretching signals in the IR spectra. 

\subsection{Ionic conductivity of confined water}

To further understand the dissociation of water molecules, we calculated the ionic conductivity ($\sigma$) of confined water at extreme P-T conditions to compare with the bulk phase. 
We obtained $\sim$0.35$\pm$0.04 $(\Omega\cdot cm)^{-1}$ at $\sim$10 GPa and 1000 K, and $\sim$9.11$\pm$2.0 $(\Omega\cdot cm)^{-1}$ at $\sim$20 GPa and 1000 K. 
Rozsa's results for the bulk water at the same P-T conditions are $\sim$1 $(\Omega\cdot cm)^{-1}$ and $\sim$10 $(\Omega\cdot cm)^{-1}$, respectively \cite{abinitio2018}.
Nanoconfinement results in a decrease in the value of $\sigma$ for water compared to its bulk phase at $\sim$10 GPa and 1000 K, suggesting that the mobility of charged ions is impeded in the graphene confined water.
It is noteworthy that water molecules are dissociated more under nanoconfinement, but the ionic conductivity actually decreases at $\sim$10 GPa and 1000 K. 
The reason is that the OH bonds indeed break more frequently, but the generated protons mostly hop between molecules and ions locally as shown in Fig. \ref{projected-O-H}(a).

\section{conclusions}

In summary, we performed AIMD simulations to compute the Raman and IR spectra of water nanoconfined by graphene at ambient and extreme P-T conditions from first principles.
We compared the calculated spectra of water under confinement and in the bulk phase, and found that the extreme spacial confinement alters the Raman stretching band and low frequency band. 
By analyzing the inter- and intramolecular contributions, we found spectroscopic evidence indicating that nanoconfinement considerably change the tetrahedral hydrogen bond network, which is typically found in bulk water.
We further investigated the impact of extreme P-T conditions on the Raman and IR spectra of confined water.
We observed an unexpected similarity in the bending mode and stretching mode intensities at around 10 GPa and 1000 K, which is attributed to the unique molecular structure of confined ionic water. 
Despite the promotion of water dissociation under nanoconfinement, our simulations suggest that proton transfer mostly occurs between nearby water molecules, resulting in lower ionic conductivity compared to that of bulk water at around 10 GPa and 1000 K. 
Additionally, we found that as the pressure increased to approximately 20 GPa while maintaining the temperature at 1000 K, confined water transformed into a superionic fluid, making it challenging to identify the stretching band of the IR spectrum.

Our study suggests that Raman and IR spectroscopy is a powerful tool to explore the structural and dynamic properties of nanoconfined water both at ambient and extreme P-T conditions. Our predicted Raman and IR spectra provide a molecular-level understanding of graphene confined water in a wide range of P-T conditions, 
which could serve as a valuable guide for future experiments. 
At ambient conditions, the stretching band and the low-frequency band in the Raman or IR spectra of water change considerably from the bulk phase to the nanoconfinement phase, which may be measured by experimentalists to study the effects of nanoconfinement.
At $\sim$10 GPa and 1000 K, the Raman spectrum of nanoconfined water has an unusual bending band related to the special structure of ionic water under confinement, which is not clearly observed in the IR spectrum, so it is recommended to conduct Raman spectroscopy measurements at this P-T condition.
It seems the overall shape of IR spectra changes more significantly than that of Raman spectra when O-H bonds break, so
it would be very informative to measure IR spectra under extreme P-T conditions; however, it is technically more difficult to conduct IR spectroscopy measurements under extreme conditions than Raman spectroscopy.
Recently, Das et al. theoretically proposed that vibrational sum frequency generation spectroscopy can be also used to study nanoconfined water \cite{das2023deciphering}, but experimental realizations are even more challenging than Raman and IR spectroscopy.

\section{Author contributions}
D.P. designed the research. 
R. H. and C. L. performed the simulations and calculations. All authors contributed to the analysis and discussion of the data and the writing of the manuscript.

\section{Competing Interests}
The authors declare no competing interests.

\section{Acknowledgements}
D.P. acknowledges support from the Croucher Foundation through the Croucher Innovation Award, Hong Kong Research Grants Council (GRF-16306621, C6021-19EF, and C6011-20GF), National Natural Science Foundation of China through Excellent Young Scientists Fund, and the Hetao Shenzhen/Hong Kong Innovation and Technology Cooperation (HZQB-KCZYB-2020083).
Part of this work was carried out using computational resources from the National Supercomputer Center in Guangzhou, China.

\bibliography{ref}

\begin{thebibliography}{10}
\expandafter\ifx\csname url\endcsname\relax
  \def\url#1{\texttt{#1}}\fi
\expandafter\ifx\csname urlprefix\endcsname\relax\def\urlprefix{URL }\fi
\providecommand{\bibinfo}[2]{#2}
\providecommand{\eprint}[2][]{\url{#2}}

\bibitem{gautam2017structure}
\bibinfo{author}{Gautam, S.~S.}, \bibinfo{author}{Ok, S.} \&
  \bibinfo{author}{Cole, D.~R.}
\newblock \bibinfo{title}{{Structure and dynamics of confined C-O-H fluids
  relevant to the subsurface: Application of magnetic resonance, neutron
  scattering, and molecular dynamics simulations}}.
\newblock \emph{\bibinfo{journal}{Front. Earth Sci.}}
  \textbf{\bibinfo{volume}{5}}, \bibinfo{pages}{43} (\bibinfo{year}{2017}).

\bibitem{Marquardt2018Structure}
\bibinfo{author}{Marquardt, K.} \& \bibinfo{author}{Faul, U.~H.}
\newblock \bibinfo{title}{{The structure and composition of olivine grain
  boundaries: 40 years of studies, status and current developments}}.
\newblock \emph{\bibinfo{journal}{Phys. Chem. Miner.}}
  \textbf{\bibinfo{volume}{45}}, \bibinfo{pages}{139--172}
  (\bibinfo{year}{2018}).

\bibitem{stolte2022nanoconfinement}
\bibinfo{author}{Stolte, N.}, \bibinfo{author}{Hou, R.} \&
  \bibinfo{author}{Pan, D.}
\newblock \bibinfo{title}{Nanoconfinement facilitates reactions of carbon
  dioxide in supercritical water}.
\newblock \emph{\bibinfo{journal}{Nat. Commun.}} \textbf{\bibinfo{volume}{13}},
  \bibinfo{pages}{5932} (\bibinfo{year}{2022}).

\bibitem{Munoz-Santiburcio2021Confinement}
\bibinfo{author}{Mu{\~{n}}oz-Santiburcio, D.} \& \bibinfo{author}{Marx, D.}
\newblock \bibinfo{title}{{Confinement-controlled aqueous chemistry within
  nanometric slit pores}}.
\newblock \emph{\bibinfo{journal}{Chem. Rev.}} \textbf{\bibinfo{volume}{121}},
  \bibinfo{pages}{6293--6320} (\bibinfo{year}{2021}).

\bibitem{lowdielectric2018}
\bibinfo{author}{Fumagalli, L.} \emph{et~al.}
\newblock \bibinfo{title}{Anomalously low dielectric constant of confined
  water}.
\newblock \emph{\bibinfo{journal}{Science}} \textbf{\bibinfo{volume}{360}},
  \bibinfo{pages}{1339--1342} (\bibinfo{year}{2018}).

\bibitem{nanoconfinement2022Ruiz}
\bibinfo{author}{Ruiz-Barragan, S.} \emph{et~al.}
\newblock \bibinfo{title}{Nanoconfinement effects on water in narrow
  graphene-based slit pores as revealed by thz spectroscopy}.
\newblock \emph{\bibinfo{journal}{Phys. Chem. Chem. Phys.}}
  \textbf{\bibinfo{volume}{24}}, \bibinfo{pages}{24734--24747}
  (\bibinfo{year}{2022}).

\bibitem{lowdielectric2022}
\bibinfo{author}{Dufils, T.} \emph{et~al.}
\newblock \bibinfo{title}{Understanding the anomalously low dielectric constant
  of confined water: an ab initio study} (\bibinfo{year}{2022}).
\newblock \eprint{arXiv:2211.14035}.

\bibitem{2Dice2016}
\bibinfo{author}{Chen, J.}, \bibinfo{author}{Schusteritsch, G.},
  \bibinfo{author}{Pickard, C.~J.}, \bibinfo{author}{Salzmann, C.~G.} \&
  \bibinfo{author}{Michaelides, A.}
\newblock \bibinfo{title}{Two dimensional ice from first principles: Structures
  and phase transitions}.
\newblock \emph{\bibinfo{journal}{Phy. Rev. Lett.}}
  \textbf{\bibinfo{volume}{116}}, \bibinfo{pages}{025501}
  (\bibinfo{year}{2016}).

\bibitem{2Dice2022}
\bibinfo{author}{Kapil, V.} \emph{et~al.}
\newblock \bibinfo{title}{The first-principles phase diagram of monolayer
  nanoconfined water}.
\newblock \emph{\bibinfo{journal}{Nature}} \textbf{\bibinfo{volume}{609}},
  \bibinfo{pages}{512--516} (\bibinfo{year}{2022}).

\bibitem{auer2008ir}
\bibinfo{author}{Auer, B.} \& \bibinfo{author}{Skinner, J.}
\newblock \bibinfo{title}{Ir and raman spectra of liquid water: Theory and
  interpretation}.
\newblock \emph{\bibinfo{journal}{J. of Chem. Phys.}}
  \textbf{\bibinfo{volume}{128}}, \bibinfo{pages}{224511}
  (\bibinfo{year}{2008}).

\bibitem{pan2020first}
\bibinfo{author}{Pan, D.} \& \bibinfo{author}{Galli, G.}
\newblock \bibinfo{title}{A first principles method to determine speciation of
  carbonates in supercritical water}.
\newblock \emph{\bibinfo{journal}{Nat. Commun.}} \textbf{\bibinfo{volume}{11}},
  \bibinfo{pages}{421} (\bibinfo{year}{2020}).

\bibitem{spectro-review2010}
\bibinfo{author}{Bakker, H.} \& \bibinfo{author}{Skinner, J.}
\newblock \bibinfo{title}{Vibrational spectroscopy as a probe of structure and
  dynamics in liquid water}.
\newblock \emph{\bibinfo{journal}{Chem. Rev.}} \textbf{\bibinfo{volume}{110}},
  \bibinfo{pages}{1498--1517} (\bibinfo{year}{2010}).

\bibitem{spectro-review2016}
\bibinfo{author}{Perakis, F.} \emph{et~al.}
\newblock \bibinfo{title}{Vibrational spectroscopy and dynamics of water}.
\newblock \emph{\bibinfo{journal}{Chem. Rev.}} \textbf{\bibinfo{volume}{116}},
  \bibinfo{pages}{7590--7607} (\bibinfo{year}{2016}).

\bibitem{spectro-review2020Seki}
\bibinfo{author}{Seki, T.} \emph{et~al.}
\newblock \bibinfo{title}{The bending mode of water: A powerful probe for
  hydrogen bond structure of aqueous systems}.
\newblock \emph{\bibinfo{journal}{J. Phys. Chem. Lett.}}
  \textbf{\bibinfo{volume}{11}}, \bibinfo{pages}{8459--8469}
  (\bibinfo{year}{2020}).

\bibitem{gygi2005ab}
\bibinfo{author}{Gygi, F.} \& \bibinfo{author}{Galli, G.}
\newblock \bibinfo{title}{Ab initio simulation in extreme conditions}.
\newblock \emph{\bibinfo{journal}{Mater. Today}} \textbf{\bibinfo{volume}{8}},
  \bibinfo{pages}{26--32} (\bibinfo{year}{2005}).

\bibitem{car1985unified}
\bibinfo{author}{Car, R.} \& \bibinfo{author}{Parrinello, M.}
\newblock \bibinfo{title}{Unified approach for molecular dynamics and
  density-functional theory}.
\newblock \emph{\bibinfo{journal}{Phys. Rev. Lett.}}
  \textbf{\bibinfo{volume}{55}}, \bibinfo{pages}{2471--2474}
  (\bibinfo{year}{1985}).

\bibitem{galli1991ab}
\bibinfo{author}{Galli, G.} \& \bibinfo{author}{Parrinello, M.}
\newblock \bibinfo{title}{{Ab-initio molecular dynamics: Principles and
  practical implementation}}.
\newblock In \bibinfo{editor}{Meyer, M.} \& \bibinfo{editor}{Pontikis, V.}
  (eds.) \emph{\bibinfo{booktitle}{Computer Simulation in Materials Science}},
  \bibinfo{pages}{283--304} (\bibinfo{publisher}{Springer},
  \bibinfo{year}{1991}).

\bibitem{marx2009ab}
\bibinfo{author}{Marx, D.} \& \bibinfo{author}{Hutter, J.}
\newblock \emph{\bibinfo{title}{{Ab Initio Molecular Dynamics: Basic Theory and
  Advanced Methods}}} (\bibinfo{publisher}{Cambridge University Press},
  \bibinfo{year}{2009}).

\bibitem{Gygi2008Architecture}
\bibinfo{author}{Gygi, F.}
\newblock \bibinfo{title}{{Architecture of Qbox: A scalable first-principles
  molecular dynamics code}}.
\newblock \emph{\bibinfo{journal}{IBM J. Res. Dev.}}
  \textbf{\bibinfo{volume}{52}}, \bibinfo{pages}{137--144}
  (\bibinfo{year}{2008}).

\bibitem{Perdew1996Generalized}
\bibinfo{author}{Perdew, J.~P.}, \bibinfo{author}{Burke, K.} \&
  \bibinfo{author}{Ernzerhof, M.}
\newblock \bibinfo{title}{{Generalized gradient approximation made simple}}.
\newblock \emph{\bibinfo{journal}{Phys. Rev. Lett.}}
  \textbf{\bibinfo{volume}{77}}, \bibinfo{pages}{3865--3868}
  (\bibinfo{year}{1996}).

\bibitem{gillan2016perspective}
\bibinfo{author}{Gillan, M.~J.}, \bibinfo{author}{Alf{\`{e}}, D.} \&
  \bibinfo{author}{Michaelides, A.}
\newblock \bibinfo{title}{{Perspective: How good is DFT for water?}}
\newblock \emph{\bibinfo{journal}{J. Chem. Phys.}}
  \textbf{\bibinfo{volume}{144}}, \bibinfo{pages}{130901}
  (\bibinfo{year}{2016}).

\bibitem{Pan2013Dielectric}
\bibinfo{author}{Pan, D.}, \bibinfo{author}{Spanu, L.},
  \bibinfo{author}{Harrison, B.}, \bibinfo{author}{Sverjensky, D.~A.} \&
  \bibinfo{author}{Galli, G.}
\newblock \bibinfo{title}{{Dielectric properties of water under extreme
  conditions and transport of carbonates in the deep Earth}}.
\newblock \emph{\bibinfo{journal}{Proc. Natl. Acad. Sci. U. S. A.}}
  \textbf{\bibinfo{volume}{110}}, \bibinfo{pages}{6646--6650}
  (\bibinfo{year}{2013}).

\bibitem{Pan2014Refractive}
\bibinfo{author}{Pan, D.}, \bibinfo{author}{Wan, Q.} \& \bibinfo{author}{Galli,
  G.}
\newblock \bibinfo{title}{{The refractive index and electronic gap of water and
  ice increase with increasing pressure}}.
\newblock \emph{\bibinfo{journal}{Nat. Commun.}} \textbf{\bibinfo{volume}{5}},
  \bibinfo{pages}{3919} (\bibinfo{year}{2014}).

\bibitem{zhang2010first}
\bibinfo{author}{Zhang, C.}, \bibinfo{author}{Donadio, D.} \&
  \bibinfo{author}{Galli, G.}
\newblock \bibinfo{title}{First-principle analysis of the ir stretching band of
  liquid water}.
\newblock \emph{\bibinfo{journal}{The Journal of Physical Chemistry Letters}}
  \textbf{\bibinfo{volume}{1}}, \bibinfo{pages}{1398--1402}
  (\bibinfo{year}{2010}).

\bibitem{quantumcal2013wan}
\bibinfo{author}{Wan, Q.}, \bibinfo{author}{Spanu, L.}, \bibinfo{author}{Galli,
  G.~A.} \& \bibinfo{author}{Gygi, F.}
\newblock \bibinfo{title}{Raman spectra of liquid water from ab initio
  molecular dynamics: vibrational signatures of charge fluctuations in the
  hydrogen bond network}.
\newblock \emph{\bibinfo{journal}{J. Chem. Theory Comput.}}
  \textbf{\bibinfo{volume}{9}}, \bibinfo{pages}{4124--4130}
  (\bibinfo{year}{2013}).

\bibitem{abinitio2018}
\bibinfo{author}{Rozsa, V.}, \bibinfo{author}{Pan, D.},
  \bibinfo{author}{Giberti, F.} \& \bibinfo{author}{Galli, G.}
\newblock \bibinfo{title}{Ab initio spectroscopy and ionic conductivity of
  water under earth mantle conditions}.
\newblock \emph{\bibinfo{journal}{Pro. Natl. Acad. Sci. U.S.A.}}
  \textbf{\bibinfo{volume}{115}}, \bibinfo{pages}{6952--6957}
  (\bibinfo{year}{2018}).

\bibitem{Brandenburg2019Physisorption}
\bibinfo{author}{Brandenburg, J.~G.} \emph{et~al.}
\newblock \bibinfo{title}{{Physisorption of water on graphene: Subchemical
  accuracy from many-body electronic structure methods}}.
\newblock \emph{\bibinfo{journal}{J. Phys. Chem. Lett.}}
  \textbf{\bibinfo{volume}{10}}, \bibinfo{pages}{358--368}
  (\bibinfo{year}{2019}).

\bibitem{PhysRevLett.89.117602}
\bibinfo{author}{Souza, I.}, \bibinfo{author}{\'I\~niguez, J.} \&
  \bibinfo{author}{Vanderbilt, D.}
\newblock \bibinfo{title}{First-principles approach to insulators in finite
  electric fields}.
\newblock \emph{\bibinfo{journal}{Phys. Rev. Lett.}}
  \textbf{\bibinfo{volume}{89}}, \bibinfo{pages}{117602}
  (\bibinfo{year}{2002}).

\bibitem{marzari2012maximally}
\bibinfo{author}{Marzari, N.}, \bibinfo{author}{Mostofi, A.~A.},
  \bibinfo{author}{Yates, J.~R.}, \bibinfo{author}{Souza, I.} \&
  \bibinfo{author}{Vanderbilt, D.}
\newblock \bibinfo{title}{Maximally localized wannier functions: Theory and
  applications}.
\newblock \emph{\bibinfo{journal}{Rev. Mod. Phys.}}
  \textbf{\bibinfo{volume}{84}}, \bibinfo{pages}{1419} (\bibinfo{year}{2012}).

\bibitem{stengel2006accurate}
\bibinfo{author}{Stengel, M.} \& \bibinfo{author}{Spaldin, N.~A.}
\newblock \bibinfo{title}{Accurate polarization within a unified wannier
  function formalism}.
\newblock \emph{\bibinfo{journal}{Phys. Rev. B}} \textbf{\bibinfo{volume}{73}},
  \bibinfo{pages}{075121} (\bibinfo{year}{2006}).

\bibitem{pan2018communication}
\bibinfo{author}{Pan, D.}, \bibinfo{author}{Govoni, M.} \&
  \bibinfo{author}{Galli, G.}
\newblock \bibinfo{title}{Communication: Dielectric properties of condensed
  systems composed of fragments}.
\newblock \emph{\bibinfo{journal}{J. Chem. Phys.}}
  \textbf{\bibinfo{volume}{149}}, \bibinfo{pages}{051101}
  (\bibinfo{year}{2018}).

\bibitem{resta2007theory}
\bibinfo{author}{Resta, R.} \& \bibinfo{author}{Vanderbilt, D.}
\newblock \bibinfo{title}{Theory of polarization: a modern approach}.
\newblock \emph{\bibinfo{journal}{Physics of ferroelectrics: a modern
  perspective}} \bibinfo{pages}{31--68} (\bibinfo{year}{2007}).

\bibitem{cavazzoni1999superionic}
\bibinfo{author}{Cavazzoni, C.} \emph{et~al.}
\newblock \bibinfo{title}{Superionic and metallic states of water and ammonia
  at giant planet conditions}.
\newblock \emph{\bibinfo{journal}{Science}} \textbf{\bibinfo{volume}{283}},
  \bibinfo{pages}{44--46} (\bibinfo{year}{1999}).

\bibitem{PhysRevLett.95.187401}
\bibinfo{author}{Sharma, M.}, \bibinfo{author}{Resta, R.} \&
  \bibinfo{author}{Car, R.}
\newblock \bibinfo{title}{Intermolecular dynamical charge fluctuations in
  water: A signature of the h-bond network}.
\newblock \emph{\bibinfo{journal}{Phys. Rev. Lett.}}
  \textbf{\bibinfo{volume}{95}}, \bibinfo{pages}{187401}
  (\bibinfo{year}{2005}).

\bibitem{goncharov2012raman}
\bibinfo{author}{Goncharov, A.~F.}
\newblock \bibinfo{title}{Raman spectroscopy at high pressures}.
\newblock \emph{\bibinfo{journal}{International Journal of Spectroscopy}}
  \textbf{\bibinfo{volume}{2012}} (\bibinfo{year}{2012}).

\bibitem{Munoz-Santiburcio2017Nanoconfinement}
\bibinfo{author}{Mu{\~{n}}oz-Santiburcio, D.} \& \bibinfo{author}{Marx, D.}
\newblock \bibinfo{title}{{Nanoconfinement in slit pores enhances water
  self-dissociation}}.
\newblock \emph{\bibinfo{journal}{Phys. Rev. Lett.}}
  \textbf{\bibinfo{volume}{119}}, \bibinfo{pages}{056002}
  (\bibinfo{year}{2017}).

\bibitem{das2023deciphering}
\bibinfo{author}{Das, B.}, \bibinfo{author}{Ruiz-Barragan, S.} \&
  \bibinfo{author}{Marx, D.}
\newblock \bibinfo{title}{Deciphering the properties of nanoconfined aqueous
  solutions by vibrational sum frequency generation spectroscopy}.
\newblock \emph{\bibinfo{journal}{J. Phys. Chem. Lett.}}
  \textbf{\bibinfo{volume}{14}}, \bibinfo{pages}{1208--1213}
  (\bibinfo{year}{2023}).

\bibitem{ohto2019accessing}
\bibinfo{author}{Ohto, T.} \emph{et~al.}
\newblock \bibinfo{title}{Accessing the accuracy of density functional theory
  through structure and dynamics of the water--air interface}.
\newblock \emph{\bibinfo{journal}{J. Phys. Chem. Lett.}}
  \textbf{\bibinfo{volume}{10}}, \bibinfo{pages}{4914--4919}
  (\bibinfo{year}{2019}).

\bibitem{Wan2015PhD}
\bibinfo{author}{Wan, Q.}
\newblock \emph{\bibinfo{title}{First Principles Simulations of Vibrational
  Spectra of Aqueous Systems}}.
\newblock Ph.D. thesis, \bibinfo{school}{University of Chicago},
  \bibinfo{address}{Chicago, IL. USA} (\bibinfo{year}{2015}).

\end{thebibliography}

\newpage

\begin{figure}
\centering
\vspace{5mm}
\includegraphics[width=0.6 \textwidth]{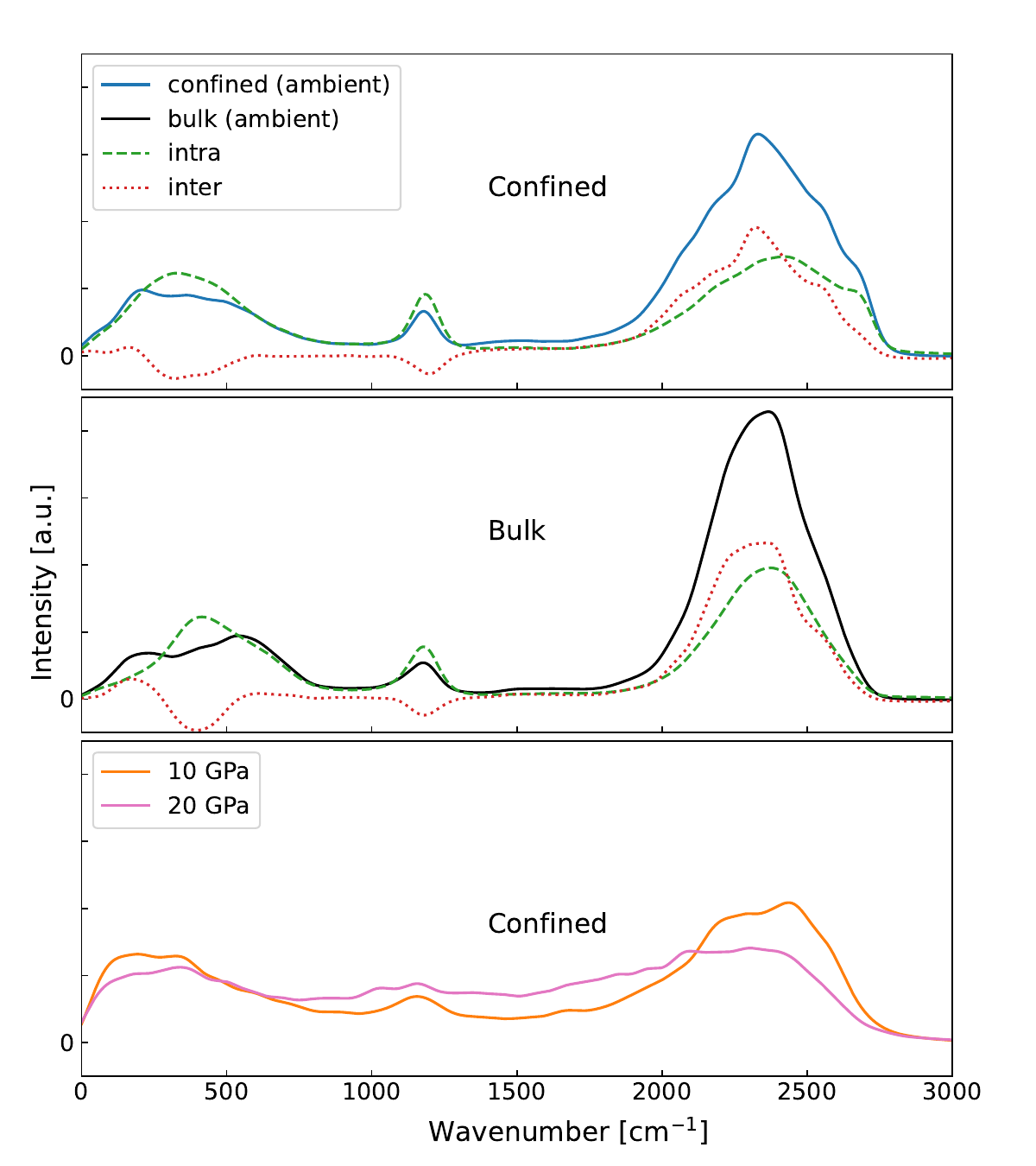}
\caption{IR spectra of graphene-confined and bulk water at ambient and extreme P-T conditions.
The upper and middle panels show the IR spectra of confined water and bulk water at ambient conditions, respectively. The intra- and intermolecular contributions are indicated by green dashed and red dotted lines, respectively. The lower panel shows the IR spectra of confined water at $\sim$10 GPa (orange lines) and $\sim$20 GPa (pink lines) with the temperature of 1000 K. The spectra of bulk water are from Ref. \cite{Wan2015PhD}. }
\label{IR}
\end{figure}

\end{document}